# A Comprehensive and Versatile Multimodal Deep Learning Approach for Predicting Diverse Properties of Advanced Materials

Shun Muroga[1], Yasuaki Miki[1], and Kenji Hata[1*]

[1]: Nano Carbon Device Research Center, National Institute of Advanced Industrial Science and Technology, Tsukuba Central 5, 1-1-1, Higashi, Tsukuba, Ibaraki, 305-8565, Japan

[*]Corresponding author. E-mail: kenji-hata@aist.go.jp

**ABSTRACT**

We present a multimodal deep learning (MDL) framework for predicting physical properties of a 10-dimensional acrylic polymer composite material by merging physical attributes and chemical data. Our MDL model comprises four modules, including three generative deep learning models for material structure characterization and a fourth model for property prediction. Our approach handles an 18-dimensional complexity, with 10 compositional inputs and 8 property outputs, successfully predicting 913,680 property data points across 114,210 composition conditions. This level of complexity is unprecedented in computational materials science, particularly for materials with undefined structures. We propose a framework to analyze the high-dimensional information space for inverse material design, demonstrating flexibility and adaptability to various materials and scales, provided sufficient data is available. This study advances future research on different materials and the development of more sophisticated models, drawing us closer to the ultimate goal of predicting all properties of all materials.

**INTRODUCTION**

The overarching goal of computational materials science is to predict all properties of all materials, and considerable efforts have been made to develop innovative approaches to achieve this objective. Classical methods, such as ab initio calculations and molecular dynamics simulations, compute atomic and electronic states based on fundamental principles of quantum and classical mechanics (**Fig. 1A.**). Although accurate, these methods are restricted to materials with simple structures, like molecules and crystals, and struggle with complex materials at larger scales. Advances in computational materials science have significantly broadened the range of materials that can be addressed, not only through improvements in classical approaches but also through new data-driven methods, including machine learning, deep learning, and generative deep learning. However, even the most advanced techniques still face challenges in predicting multiple physical properties of conventional composites like plastics, metal alloys, and rubbers, commonly used in everyday life.

One example of advanced classical simulations is high-throughput simulations, which accelerate ab initio calculations using efficient algorithms and advanced computational resources to calculate electronic states for billions of atoms and various physical properties of polymers. However, it should be noted that the materials that can be modeled are still limited to the nanoscale (*1–3*). Another example is multi-scale simulation, a computational method that combines simulations at multiple scales into a comprehensive model to accurately predict the behavior of complex materials at micro or larger scales (*4–6*). Although this approach is effective in modeling large materials, its ability to calculate diverse properties is often limited due to customization of the integrated model for specific properties. While these approaches represent significant progress, even the most advanced model has limitations in predicting a variety of properties for complex materials at large scales.

Data science, rapidly progressing and fundamentally distinct from classical methodologies, extracts highly correlated structural features of materials with desired properties and predicts material properties through structure-property correlation analysis (*7, 8*). As long as structural features are quantified and correlations with properties are established, any material or property can be modeled. Consequently, machine learning has made a significant impact on computational materials science. Molecular descriptors have been developed to quantify molecular structural features, including elemental composition, chemical bond types, adjacency relationships, and electronic interactions. These descriptors provide a standardized format for representing molecular structures, enabling researchers to explore and identify molecules with superior properties through open databases linking molecular descriptors to physical properties. For example, high-throughput screening has been reported to find optimal molecules for organic thin-film solar cells among 1,700 candidates (*9*), as well as the development of high-performance polymer gas separation membranes surpassing theoretical limits for separating carbon dioxide and methane (*10*). Additionally, artificial intelligence has been trained with ab initio calculation results to create neural network potentials that can predict atom interaction potentials without needing first-principles calculations. Instances of this approach include fast calculations of molecular diffusion in metal-organic frameworks (*11*) and electrical conductivity in metal nanowires (*12*). By employing deep learning, it becomes possible to accurately predict physical properties with highly nonlinear relationships to molecular structures. For example, there are reports of predicting the synthesizability of inorganic materials by inputting their atomic configurations into deep learning models (*13*).

Traditional data science relies on quantification of material structures and is not applicable to materials with undefined structures. In fact, most materials used in daily life, such as rubber, resin, and wood, have undefined structures, with very few data science research examples. Humans can recognize and reproduce unquantifiable features of individuals, like unique facial appearances, through art. Recently, deep learning techniques have been developed to mimic this human cognitive ability and generate images of various objects (known as generative deep learning), with potential applications in computational materials science being explored. For example, generative deep learning models have been reported to generate structures of molecules (*14*, *15*), inorganic materials (*16–18*), and crystals (*19–21*) with atomic arrangements closely resembling their real counterparts. Moreover, generative deep learning models can generate images of complex and large-scale materials that cannot be quantified in simple atomic configurations. For instance, Banko et al. demonstrated the generation of structural images of scanning electron microscopy (SEM) corresponding to the sputtering conditions of metal films with different compositions and the creation of phase diagrams using a generative deep learning model(*18*). Furthermore, Yang et al. employed a generative deep learning model to create stress distribution maps in composite materials (*22*). In our previous study, we developed a deep learning model capable of generating SEM images of films made of binary or ternary mixtures of CNTs (1716 compositions) and predicted their electrical conductivity and specific surface area based on these images (*23*). Although this result demonstrated the feasibility of using deep learning to predict physical properties from generated structural images, the range of properties that can be predicted from SEM images was found to be limited. In summary, while significant advances have been made by data science, predicting diverse properties for complex materials remains a major challenge in the field of computational materials science.

# RESULTS
## Representations of Physical and Chemical Structures

Given comprehensive material characterization, the potential for predicting its properties is limitless. Building upon this idea, we have developed a multimodal deep learning (MDL) model that leverages multiple sources of complementary information as inputs, to achieve optimal performance in forecasting properties of materials with intricate structures (**Fig. 1B.**). This approach resembles the human cognitive process, where information from different sensory modalities, such as sight and hearing, is integrated to gain a comprehensive understanding of the environment. Due to its superior recognition capabilities, MDL has gained significant attention, especially in research fields centered around human activity (*24*). For instance, information from brain wave signals and eye movements has been integrated to distinguish positive, negative, and neutral human emotions for emotion analysis (*25*). Moreover, magnetic resonance images and psychological test results have been combined to diagnose the symptoms of human dementia (*26*). Although various MDL models have been developed in the literature, their application to materials research remains unexplored.

Our MDL model consists of four separate modules: three independent generative deep learning models expressing different aspects of material structure are connected in parallel to a single deep learning model, which predicts a set of material properties as an output. This model possesses several key features. Firstly, three generative deep learning models were designed to provide not only physical features but also complementary chemical information about the material, allowing for a comprehensive characterization of its structure. Specifically, we employed Generative Adversarial Networks (GAN), a deep learning technique, to generate optical microscope images representing the material's physical structures, and spectra from infrared absorption (IR) and Raman spectroscopy to capture its chemical information. GANs mimic the human cognitive process of recognizing features of objects with undefined structures, such as human faces, and can generate corresponding images (*27, 28*). The utilization of GANs enabled us to model the structure of complex materials, such as plastic composites containing additives and fillers, whose structure cannot be numerically defined. Secondly, the fourth deep learning model was developed to integrate the physical structure and chemical information of the material generated by the GANs, which were presented in different formats, while preserving their characteristic features. This model enabled the prediction of a wide array of physical properties, a task that would have been challenging to achieve using conventional machine learning techniques.

Our approach, which incorporates the key features described above, has the potential to rapidly predict a diverse range of material properties for a wide range of compositions entirely through computational means, without the need for any experiments beyond those required to train the models. More precisely, we developed an MDL model capable of forecasting eight distinct material properties, namely, Young's modulus, tensile strength, fracture strain, glass transition temperature (Tg), density, electrical resistivity, storage modulus, and loss tangent, for a 10-dimensional composition of a typical acrylic polymer composite material composed of five matrices (AH, EH, HEMA, BZ, Ebe), two additives (TMP, MTNR), and three fillers (alumina particles, carbon fibers, carbon nanotubes).With this MDL model, we predicted 913,680 properties across 114,210 composition conditions. Our study has achieved a remarkable level of complexity in dimensionality, featuring 18 dimensions consisting of 10 inputs and 8 outputs, and in the number of data points generated, which is 913,680. This is an unprecedented achievement in the field of computational materials science, and its significance is amplified by the fact that it was realized for a material with an undefined structure.

We present the development of three generative adversarial networks (GANs) designed to generate the physical and chemical structures of composites based on their compositions. Three GAN models were created using a 10-dimensional composition input to produce optical microscope images, infrared (IR) spectra, and Raman spectra of the composites, respectively. GANs, advanced deep learning techniques, create high-fidelity data through the competition of two artificial intelligence models. One model generates the data, while the other evaluates the authenticity of the generated data (**Fig. 2A.**). A substantial amount of experimental data is required for this training process. In this study, 80 samples (listed in **Table S1.**) were fabricated with varying proportions of matrix monomer, additive, and filler, maintaining constant process parameters such as dispersion and curing. The samples were characterized by optical microscopy, IR, and Raman spectroscopy to obtain training data. For the optical microscope GAN (OM-GAN), 80,000 pixel-segments (128x128) were extracted from experimental optical microscope images taken at 100x magnification. For the IR and Raman spectra GANs (IR-GAN and Raman-GAN), data were preprocessed by selecting 1024 data points below the 1800 cm-1 range to include crucial absorption peaks and normalized to a range of 0-1. These preprocessed spectra were augmented 128-fold by adding Gaussian noise to each spectrum, resulting in 10,240 training spectra. To assess GAN training effectiveness, the Fréchet Inception Distance (FID) score (*29*) was calculated to quantify the similarity between generated and training data. The FID scores of OM-GAN, IR-GAN, and Raman-GAN steadily decreased and reached a stable value, indicating successful GAN training (**Fig. 2B.**).

A series of tests were conducted to evaluate the fidelity of the images and spectra generated by the trained GANs. First, the optical microscope images (**Fig. 2C.**), IR spectra (**Fig. 2D.**), and Raman spectra (**Fig. 2E.**) generated by the trained GANs exhibited high similarity to the corresponding actual images and spectra of the same composition. Second, the ability of GANs to accurately create images and spectra representing various compositions was verified. **Fig. 2F.** displays 64 optical microscope images generated by OM-GAN, arranged in an 8x8 grid, with varying concentrations of alumina particles. As the concentration increased, the number of white alumina particles in the images also increased, indicating that the GAN could generate images corresponding to different levels of concentration. The structural diversity of the 8x8 OM-GAM images created also demonstrates success in preventing mode collapse, a common problem in GAN training where the generator network produces similar or identical images. When the concentrations of matrix monomers AH and Ebe were varied, the IR spectra generated by IR-GAN showed significantly different shapes (**Fig. 2G.**) and were not simple superimpositions of individual monomer spectra (**Figs. S1 and S2.**). The spectral shape changes before and after crosslinking are exemplified by the disappearance of the C=C (1650 cm$^{-1}$) due to the bond cleavage reaction, shifts of the C=O (1720 cm$^{-1}$), and the aromatic C-H (1200 cm$^{-1}$) induced by the interactions in the crosslinked structures (**Fig. S3.**). Such IR GAN spectra include not only the molecular structures of the monomers but also the complex states of the crosslinked structures, which we believe is important for predicting the physical properties of the composite.

**Multimodal Model Integrating Multiple Physical and Chemical Structures.**
A deep learning model was constructed to predict eight physical properties of composite materials from optical microscope images, IR, and Raman spectra generated by GAN. To combine data in different formats, including two-dimensional images and one-dimensional spectra, the data were processed by convolutional and dense layers to extract their features and subsequently integrated by concatenate layers, resulting in a single one-dimensional vector (see **Fig. 3A.**) from which the physical properties were predicted.

Our model predicted physical properties for 114,210 compositions, despite being trained on experimental data from only 80 composite compositions. In such a situation, constructing a reliable model requires careful consideration of how the experimental data is treated. Typically, to determine the optimal training conditions, the experimental dataset is split into training and test data (hold-out method). However, when the experimental data is limited, this approach may lead to an unreliable or unstable model due to sampling bias. Thus, in this study, we generated 16 different sets of training and test data with a fixed 7:3 ratio and trained 16 models with these sets (cross-validation method). Subsequently, we calculated the mean absolute error (MAE) between the predicted and actual values of eight physical properties for each of the 16 sets of training and test data, and used the average MAE as an indicator of the model performance. The number of training epochs, a critical factor that determines model performance, was optimized by monitoring the average MAE. The trend of the MAE as a function of the number of training epochs exhibits an initial monotonic decrease for both the training and test data (**Fig. 3B.**), while the MAE of the test data increased after 400 epochs. Based on this result, the final model was trained for 400 epochs with all available experimental data. The relationship between the prediction of the final model and the experimental values of eight physical properties for all compositions is presented in **Fig. 3C.** The plots for Young's modulus, tensile strength, elongation at break, storage modulus, loss tangent, Tg, density, and electrical resistivity are concentrated near the diagonal, indicating that the constructed final model has high predictive performance for all physical properties.

**Overview of High-dimensional Information Space of Predictions.**
The MDL model's predictions constitute an information space of unprecedented size in the field of computational materials science, featuring 18 dimensions of composition and properties and 913,680 data points. Given our limited ability to visualize beyond three dimensions, comprehending the entire structure of this information space is challenging for humans. In order to address this issue, a series of 'maps' displaying property–composition relationships were created by conventional data visualization techniques. We selected four types of matrix monomers (AH, EH, HEMA, BZ) out of five, while keeping the amounts of additives and fillers constant. Since the proportions of the four types of monomers sum up to 100%, their four-dimensional compositional information could be displayed by a 3D tetrahedral coordinate system. A specific property was selected, and its predicted values were visualized by a color gradient in the tetrahedral coordinate system, creating a property-composition 'map' that represented a 5-dimensional subspace of the information space.

Property-composition 'maps' for seven properties are presented in **Fig. 4.**, which collectively visualize an 11-dimensional subspace of the information space. The diverse topographies of the 'maps' reflect the sophisticated correlations between composition and properties. Specifically, the observed rise in the elastic modulus with increasing AH content is attributed to AH's rigidity, which enhances the crosslinking density of the composite. Moreover, the monofunctional monomers, HEMA and BZ, both contribute to the increase in elastic modulus, but the extent of their impact on the elastic modulus varies. On the other hand, the incorporation of EH, a flexible monomer with a low glass transition temperature, into the polymer composite results in an increase in elongation at break and loss tangent, suggesting an increase in the material's viscosity. In addition, by focusing on the general trends among multiple properties, a negative correlation can be identified between elongation at break and the elastic modulus, strength, and glass transition temperature.

The observed correlation trends between composition and properties in the 'maps' are consistent with the expected trends based on domain knowledge of the material, thus demonstrating the reliability of the model's predictions. We would like to note that even with the 7 "maps" shown in **Fig. 4.**, only a small fraction of the 18-dimensional information space has been explored. With respect to composition, no variations in additives, fillers, or other monomers have been made, and correlations between properties that are important for practical applications have not been directly visualized, highlighting the difficulty of understanding a high-dimensional information space.

**Characterization of High-dimensional Information Space of Properties.**
In the following sections, we present our approach for understanding the entire structure of the 18-dimensional information space generated by the MDL model. As a first step, to overview the entire information space, 153 pair plots (see **Fig. 5.**) were created, consisting of both scatter plots and kernel density distributions, with the x and y axes selected from the 18 dimensions related to composition and properties. With each panel containing 114,210 data points, **Fig. 5.** displays a total of 17,474,130 points, offering a bird's-eye view of all possible correlations between the composition and properties of the composite. The shapes of the plots exhibit a high degree of diversity, highlighting the extremely complex relationships between the composition and properties of the composite. To gain a deeper insight into the overall complex correlations, we classified the 153 panels into three categories: composition-composition, composition-property, and property-property plots, and focused on analyzing the most important property-property plots. By examining these plots, we can identify trends and correlations between different properties and composition variables, allowing for a more comprehensive understanding of the material's behavior. This analysis can serve as a basis for further optimization and exploration of the composite's design, leading to the development of materials with tailored properties for specific applications.

To elucidate the relationships between physical properties, we calculated similarities between properties by representing each as a 114,210-dimensional vector, with predicted values as elements. The similarity between physical properties was quantified using the inner product (cosine distance) between vectors. Consequently, physical properties with high similarities were grouped together in the dendrogram (**Fig. 6A.**), and the eight physical properties were classified into three groups. Group 1 encompasses loss tangent and elongation at break, mechanical properties indicative of material flexibility. Group 2 comprises storage modulus, Young's modulus, and tensile strength, mechanical properties representing material hardness. Lastly, Group 3 includes properties with weaker correlations to the mechanical properties in Groups 1 and 2, such as Tg, density, and electrical resistivity. Flexibility and hardness are essential mechanical properties of materials and typically exhibit a strong inverse correlation in various materials, including resins, plastics, rubbers, metals, and ceramics. This inverse correlation arises from fundamental physical principles. Specifically, hardness is achieved by strongly bonding rigid materials and restricting deformation, while flexibility is attained by loosely bonding soft materials and permitting free deformation. Such intrinsic inverse correlation between physical properties is evident in a wide range of materials, including battery power and energy (*30*), as well as the Seebeck coefficient and electrical conductivity of thermoelectric materials (*31*). We propose categorizing material physical properties based on intrinsic, fundamental physical principles, using dendrograms as a guide.

Subsequently, property-property plots (also known as Ashby plots) were classified into six zones by segmenting them with dashed lines among the three property groups, as illustrated in **Fig. 6B.** Observations indicate that Ashby plot shapes are highly similar within the same zone but differ significantly across regions, suggesting that Ashby plots can be classified into several types based on their respective zones. Based on their shapes, Ashby plots were classified into four typical types: monotonic increase (light blue), monotonic decrease (red), up-and-down (orange), and kinked (green), as shown in **Fig. 6C.** Plots of the same type were enclosed in a colored square, clearly demonstrating a correlation between zones and types. Observations indicate that Ashby plots exhibit similar shapes within the same zone. We interpret that an underlying physical mechanism governs the correlation between different properties. For instance, the storage elastic modulus, Young's modulus, and tensile strength within Group 2 all exhibit a consistent monotonic increase trend, as these properties are primarily determined by the material's hardness. Moreover, all Ashby plots in the Group 1-2 zone reveal a complementary relationship between the two properties: as one property increases, the other decreases. This highlights the inherent trade-off between material hardness and softness and illustrates the challenge of simultaneously increasing both properties. The discontinuous trends observed in types such as up-and-down (orange) and kinked (green) cannot be solely explained by the material's physical mechanism, unlike the monotonic increase/decrease scenarios. Instead, these trends are associated with significant variations in composition, as described later. As demonstrated, methodologies such as grouping multiple physical properties based on similarities and classifying trends in Ashby plots are extremely useful for understanding complex correlation relationships between physical properties.

**<u>Inverse Materials Design.</u>**

In the previous section, we presented methodologies for analyzing relationships between physical properties. In this section, we will delve further and introduce our proposed analytical approach for investigating relationships between physical properties and composition. Specifically, we will focus on examples and methodologies that extract valuable compositional information from the 18-dimensional information space, enabling inverse material design. Inverse material design involves designing new materials by starting with the desired properties and working backward to determine the material's composition and structure. Only a portion of the physical property predictions from a model is meaningful for material design. This is because, in material design, only compositions that exhibit an optimal solution in terms of the desired properties are considered as candidates. To provide examples, one representative Ashby plot was chosen from each of the four types. In each of the four panels, there are 114,210 predicted results shown in blue, from which the set of non-dominated solutions, i.e., points at the upper boundary, were selected and indicated by yellow stars. The set of non-dominated solutions refers to a collection of solutions in a multi-objective optimization problem that are not dominated by any other solutions. In most cases, the set of non-dominated solutions would be the candidates for material design, and thus, as an examples. we have displayed the corresponding composition diagram of the solutions in the four panels of **Fig. 7.**

By analyzing the set of non-dominated solutions in the Ashby plot and the corresponding composition diagrams, we can track the changes in optimal composition that occur with variations in material properties. For instance, examining the increasing trends of tensile strength and Young's modulus (**Fig. 7A.**), we observed a gradual shift in the optimal solution composition, with transitions among EH, BZ, AH, and HEMA. This shift aligns with monomer stiffness. In **Fig. 7B.**, the optimal solution composition for storage modulus and loss tangent exhibited a distinct transition compared to **Fig. 7A.**, as EH and HEMA's monofunctional structure reduces crosslinking density, increasing the loss tangent. These monomers were identified as optimal solutions. The Pareto frontier represents the set of optimal solutions for contrasting material property relationships and is crucial in material design (*3, 32–34*). Conversely, discontinuous and sharp phase transitions were found in the up-and-down (**Fig. 7C.**) and kinked (**Fig. 7D.**) Ashby plots. In **Fig. 7C.**, yellow EH increased in a Tg region, interrupting the optimal composition's gradual change, attributed to distinct flexibility mechanisms involving low crosslinking density and low Tg molecular structure. In the kinked Ashby plot, a sharp transition from purple Ebe to green BZ occurred above a certain Tg (**Fig. 7D.**), as higher levels of green BZ are optimal for enhancing Young's modulus in the Tg range without purple Ebe. We would like to highlight that the properties-composites combination obtained through this approach represent potential candidates for inverse material design, a methodology that involves designing new materials by specifying the desired properties and then determining the material's composition and structure.

**DISCUSSION**

We have developed a rational MDL model that demonstrates a significant advancement in the field of computational materials science, as it addresses the challenge of predicting diverse properties for complex materials. By leveraging multiple sources of complementary information and integrating them into a single model, we have created a powerful tool that can effectively predict a wide array of physical properties for a broad range of materials. Our strategy relies on employing generative deep learning models to generate physical structures (optical microscope images) and chemical information (IR and Raman spectra) in accordance with the material composition, utilizing GANs and integrating them to forecast a multitude of physical properties. This approach is especially invaluable in circumstances where conventional methodologies prove ineffective due to the structural intricacy. It is applicable to an extensive assortment of materials, such as metals, polymers, and ceramics, and can be employed across systems of diverse sizes, ranging from micro- to macroscopic scales.

Our MDL model predicted over 900,000 properties in 114,210 compositional conditions, resulting in an unprecedented level of complexity with 18 dimensions, including 10 compositional inputs and 8 property outputs. The approach can be easily expanded to represent more complex materials or devices by interconnecting MDL models, where each model represents a fabrication process step, and the integrated MDL represents the entire fabrication process for a system or device made from various materials.

We present an analytical approach to simplify the understanding of high-dimensional information spaces generated by advanced models, which may be too complex for humans to fully comprehend. Our method involves grouping physical properties based on similarities using dendrograms and classifying trends in Ashby plots to identify correlations. We further analyze categorized Ashby plots by extracting non-dominated solutions and displaying corresponding compositions, which serve as candidates for inverse material design. This design strategy starts with desired properties and determines the material's composition and structure. By examining these plots, we identify trends and correlations between various properties and composition variables, enabling a comprehensive understanding of material behavior. Our analysis lays the foundation for further optimization and exploration of composite designs, leading to the development of materials with tailored properties for specific applications. In summary, the successful application of MDL to materials research opens the door for future studies and the development of even more advanced models, ultimately bringing us closer to the ultimate objective of computational materials science: the prediction of every property of all materials.

## MATERIALS AND METHODS
### Fabrication of Polymer Composites

In this study, acrylate polymer composite materials were fabricated using a variety of matrices, additives, and fillers. Specifically, five different acrylate monomers were used as matrices: phenyl glycidyl ether acrylate hexamethylene diisocyanate urethane prepolymer (AH-600, KYOEISHA CHEMICAL Co., LTD), 2-ethylhexyl methacrylate (Tokyo Chemical Industry Co., Ltd.), 2-hydroxyethyl methacrylate (Tokyo Chemical Industry Co., Ltd.), benzyl acrylate (Lightester BZ, KYOEISHA CHEMICAL Co., LTD), bifunctional urethane acrylate (EBECRYL 230, DAICEL-ALLNEX LTD.). These matrix monomers were referred to as AH, EH, HEMA, BZ, and Ebe for short. The molecular structures of the matrix monomers used in the study are illustrated in **Fig. S1**. Additionally, two different additives were utilized: trimethylolpropane trimethacrylate (Lightester TMP, KYOEISHA CHEMICAL Co., LTD) as a curing additive, and 1,3,5-tris (2-(3-sulphanylbutanoyloxy)ethyl)-1,3,5-triazinans-2,4,6-trion (Karenz MT-NR1, Showa Denko K. K.) as a chain transfer agent, which were abbreviated as TMP and MTNR respectively. In order to initiate the curing process, a fixed amount of t-amyl peroxy-2-ethyl hexanoate (Luperox 575, from ARKEMA Yoshitomi, Ltd.) was added to the materials at a concentration of 1 phr (parts per hundred resin). The molecular structures of the additives used in this study are illustrated in **Fig. S1.** In this study, three different types of fillers were used to fabricate acrylate polymer composite materials. These fillers included alumina (ALUNABEADS CB-A20S, Showa Denko K. K.) with a nominal particle diameter of 21 μm, used as a spherical filler; chopped carbon fiber CF (DIALEAD K223HE, Mitsubishi Chemical) with a nominal diameter of 11 μm and no sizing agent, used as a rigid fibrous filler; and single-walled carbon nanotube CNT (ZEONANO SG101, Zeon Corp.) used as a flexible fibrous filler. The fillers were added to the materials in specific concentrations, with alumina and CF added in amounts less than 30 phr, and CNT added in amounts less than 0.5 phr. The optical microscope images of the raw fillers are depicted in **Fig. S4.** These materials were mixed using a rotational mixer at a high speed of 10000 rpm for 20 minutes. Then, the resulting mixture was cured at a temperature of 100 °C for 60 minutes to produce a polymer composite sheet that was 1mm thick. A total of 80 samples with varying compositions were created and utilized for this study (**Table S1.**)

### Characterization

The properties of the composite materials were analyzed by acquiring optical microscope images using a digital microscope (VHX-7000 from Keyence Co., Ltd.) at a magnification of 100x. IR spectra were obtained using a Fourier-transform infrared spectrometer (FT-IR-4600 from JASCO Co., Ltd.) with a resolution of 4 cm$^{-1}$, using an attenuated total reflection prism made of germanium. Raman spectra were also obtained using DXR2 SmartRaman (from Thermo Fisher Scientific K.K.) at an excitation wavelength of 532 nm. The mechanical properties of the polymer composites, including Young's modulus, tensile strength, and elongation at break, were evaluated using a universal mechanical tester, the EZ-LX (Shimadzu Co., Ltd.). The tests were conducted at a crosshead speed of 5 mm/min. The rheological properties of the polymer composites were assessed using a dynamic mechanical analysis (DMA) using RSA3 (TA Instruments). The DMA measurements were performed with a heating rate of 10°C/min, a strain of 0.5%, and a frequency of 1 Hz. The storage modulus and loss tangent of the polymer composites were evaluated at room temperature using dynamic mechanical analysis (DMA). The glass transition temperature was determined by identifying the point at which the loss tangent reaches its maximum value in the DMA profiles. The electrical properties of the polymer composites were evaluated using LORESTA-GP MCP-T610 (Mitsubishi Chemical Analytech Co., Ltd.) to measure the surface resistivity. Distributions of eight physical properties of the polymer composites are shown in **Fig. S5.**

**Computational Environments**

Our computational environments in this study are as follows: python/3.6/3.6.12, cuda/10.0/10.0.130.1, cudnn/7.4/7.4.2, tensorflow-gpu/1.14.0, scikit-learn/0.23.2. For deep learning calculation, we used a computation node including the CPU of Intel Xeon Gold 6148 and the GPU of NVIDIA V100 in AI Bridging Cloud Infrastructure (ABCI) provided by National Institute of Advanced Industrial Science and Technology (AIST).

**Data Preprocessing**

The data preprocessing involves dividing the optical microscope images into smaller pieces of 128x128 pixels, with an interval of 64 pixels, and obtaining 1,000 images per composition. These images were then used to train the generator and discriminator. The IR and Raman spectra were preprocessed by selecting 1024 data points in the wavenumber range of 1800 $cm^{-1}$ and below, and normalizing them between 0-1. Gaussian noise with a mean of 0 and a standard deviation of 0.001 was added, and the data was augmented by a factor of 128. The IR and Raman spectra were preprocessed by extracting 1024 data points in the wavenumber range below 1800 $cm^{-1}$ and normalizing them in the range of 0 to 1. Gaussian noise with a mean of 0 and a standard deviation of 0.001 was added, and the data was augmented by a factor of 128. The data was then transformed into a two-dimensional array and trained on the generator and discriminator. For the regression using the integrated multimodal model, the eight target properties of the polymer composites were standardized by the mean and standard deviation of each property.

**Generative Models using Generative Adversarial Networks**

In this study, BigGAN (*28*) was employed to build three GANs: OM-GAN, IR-GAN, Raman-GAN to generate the optical microscope images and IR and Raman spectra. BigGAN is one of the conditional GANs designed for high fidelity data generation including innovative approaches, such as truncation trick, conditional batch-normalization, and orthogonal regularization. The specific architectures employed were the 128x128 model for the OM-GAN and the 32x32 model for the IR-GAN and Raman-GAN. We constructed the BigGAN models for generating images and spectra based on open-source code (*35*). Example trained GAN models in our previous study (*23*) can be accessed via a link (*36*). Detailed architectural diagrams for the generator and discriminator of the 128x128 model can be found in **Fig. S6.** and **Fig. S7.** For the generator (**Fig. S6.**), the inputs consist of latent values and a vector representing the composition, and the output is generated data with a resolution of 128x128. The generator utilizes ResBlocks which include conditional batch normalization (*37*), rectified linear unit (ReLU) activation (*38*), convolution, and upsampling, along with a skip connection (*39*) for upsampling. For the discriminator (**Fig. S7.**), the inputs are the data with a resolution of 128x128 and a composition vector, and the output is a logit. The discriminator also uses ResBlocks which include conditional batch normalization, ReLU activation, convolution, and downsampling, along with a skip connection for downsampling and convolution. The architectural designs of 32x32 BigGAN for IR-GAN and Raman-GAN are outlined in **Fig. S8.** and **Fig. S9.** The variations in the model architecture include the number of ResBlocks and the number of parameters. The training method for three GANs are illustrated in **Fig. S10.** The Hinge loss, calculated from the logits of both reference and generated data, is used as the primary loss function for the generator and discriminator. Additionally, an orthogonal regularization term is included in the loss calculation to optimize the weights of the generator and discriminator. Both the generator and discriminator weights are optimized using the Adam optimizer (*40*), with a batch size of 64 and 128 latent variables. The weights of the models were optimized using Adam optimizers with a learning rate of 0.0001 for the discriminator and 0.0004 for the generator. The values for $\beta 1$ and $\beta 2$ were set at 0 and 0.9, respectively, for both the generator and discriminator. To assess the fidelity of the data generation with the trained GANs, Fréchet inception distance (FID) scores (*29*) were calculated. FID is a metric that compares the distribution of the generated data to that of the reference data using activation vectors from an Inception v3 model (*41*) trained on ImageNet dataset (*42*). A lower FID score indicates the higher fidelity of the data generation, as it suggests a smaller difference between the generated and reference data. The FID scores for the GAN models of the optical microscope images, IR and Raman spectra were calculated using open-source code (*43*) modified to be compatible with our computational environment.

**Multimodal Model with Multiple Inputs of Physical and Chemical Structure Information**

A multimodal deep learning model was developed as shown in Fig. 3, with the inputs such as a 128x128 size optical microscope image, IR spectrum with 1024 data points, Raman spectrum with 1024 data points, and 10 different compositions, and the outputs of right physical properties of the polymer composite. Our strategy for constructing the integrated multimodal deep learning model involved dividing the model into five sections: layers for processing the optical microscope images, layers for processing infrared spectra, layers for processing Raman spectra, and layers for processing composition data, followed by layers for integrating the information. The convolutional layer was used to handle the two-dimensional data of the optical microscope images, while the dense layer was used for the one-dimensional data of the spectra. To ensure stable training, we also included batch normalization (*44*) and dropout (*45*) layers in the model architecture. The weights of the model were optimized using the Adam optimizer with a learning rate of 0.001, $\beta 1$ of 0.9, and $\beta 2$ of 0.999. To eliminate the impact of sample selection on the training and test data sets, the sampling process was randomly repeated 16 times with a fixed ratio of 7:3 for training and test (cross-validation method). Average accuracies of 16 models were calculated to reduce the influence of sampling bias. Specifically, all hyperparameters of the model include the number of convolutional or dense layers and their neurons. However, as optimal values can vary depending on the dataset and the number of possible combinations is vast, we adjusted these numbers roughly and chose balanced values for this study. In the stage of virtual screening of materials with arbitrary compositions, 256 images, IR spectra, and Raman spectra were generated by GAN models in four batches per composition, and the properties were predicted.

**Acknowledgments:** We appreciate Dr. Toshiya Okazaki, Dr. Don N. Futaba, Dr. Hiroshi Morita, Dr. Takashi Honda, Mrs. Kaori Tatsumi, Mrs. Megumi Terauchi, Mrs. Maiko Nihei, and Mr. Tomohisa Hayashida, for their supports. We also appreciate Daicel allnex Corporation for providing the materials. Computational resource of AI Bridging Cloud Infrastructure (ABCI) provided by National Institute of Advanced Industrial Science and Technology (AIST) was used. **Funding**: This work was supported by a project (JPNP16010) commissioned by the New Energy and Industrial Technology Development Organization (NEDO). **Author Contributions**: CRediT authorship statements are as follows, Conceptualization: S.M. and K.H.(leading), Methodology: S.M., Software: S.M., Formal Analysis: S.M., Investigation: S.M. and Y. M., Visualization: S.M. and K. H., Writing – Original Draft: S.M. and K.H., Supervision: K.H., Project administration: K.H. Specifically, K.H. conceived the general idea of MDL in this work, led, supervised, and managed the entire project, and prepared figures and main text of this paper. S.M. realized MDL in terms of materials, characterization, calculation, carried out all experimental and computational work, developed the MDL model, and wrote all programs including GANs, MDL, deep learning, statistical analysis, data visualizations, and prepared figures, draft of main text, supplementary materials of this paper. Y. M. contributed the fabrication of acrylic polymer composites and investigated the properties. **Competing Interest**: The authors declare no competing interests. **Data and materials availability**: All data needed to evaluate the conclusions in this paper are present in the paper and/or the Supplementary Materials. Additional supporting data generated during the present study are available from the corresponding author upon reasonable request.

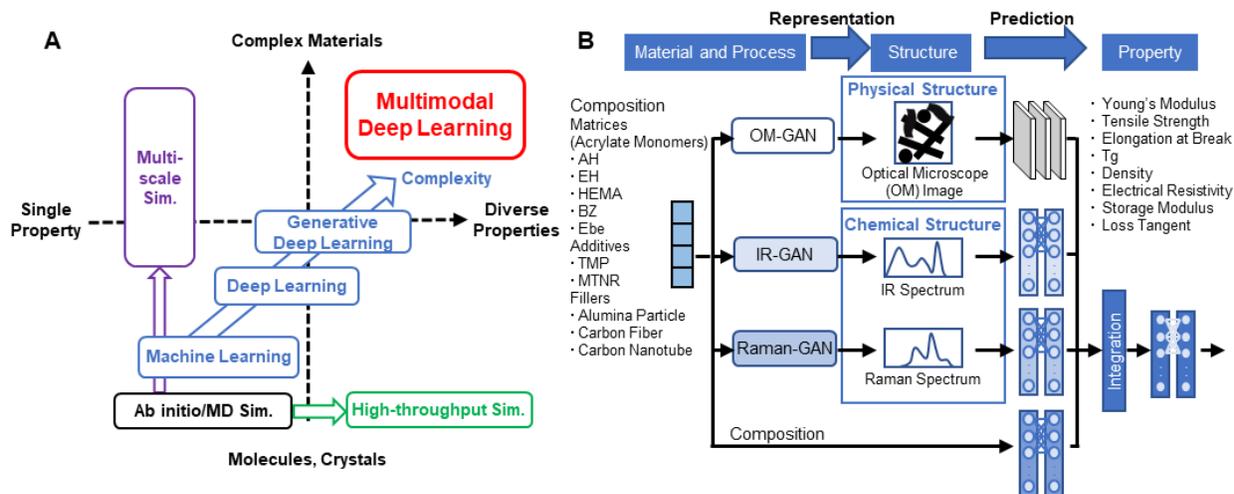

**Fig. 1. Computational Materials Science.** (A) Classification of traditional simulations, machine learning, deep learning, and multimodal deep learning for computational materials science. The vertical axis represents the complexity of materials, and the horizontal axis represents the complexity of properties. (B) The proposed multimodal deep learning integrating multiple physical and chemical structure information, enabling inferences about various properties of complex materials.

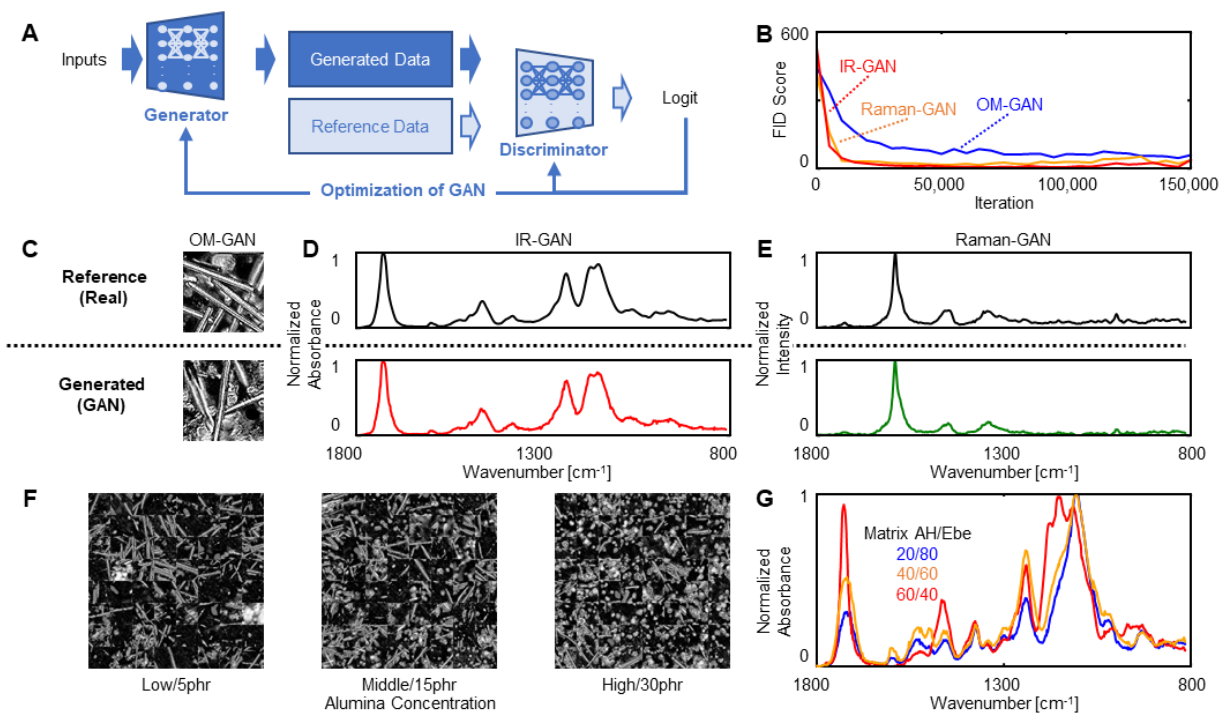

**Fig. 2. Representations of Physical and Chemical Structures using Generative Deep Learning.**
(A) A diagram of the generative adversarial network (GAN), a kind of generative deep learning. (B) A graph showing the progression of training GAN models, as measured by Fréchet inception distance (FID) scores. Real and AI-generated (C) optical microscope images, (D) infrared spectra, and (E) Raman spectra. The trained GAN models were used to generate (F) optical microscope images with varying filler content, and (G) infrared spectra with different matrix compositions.

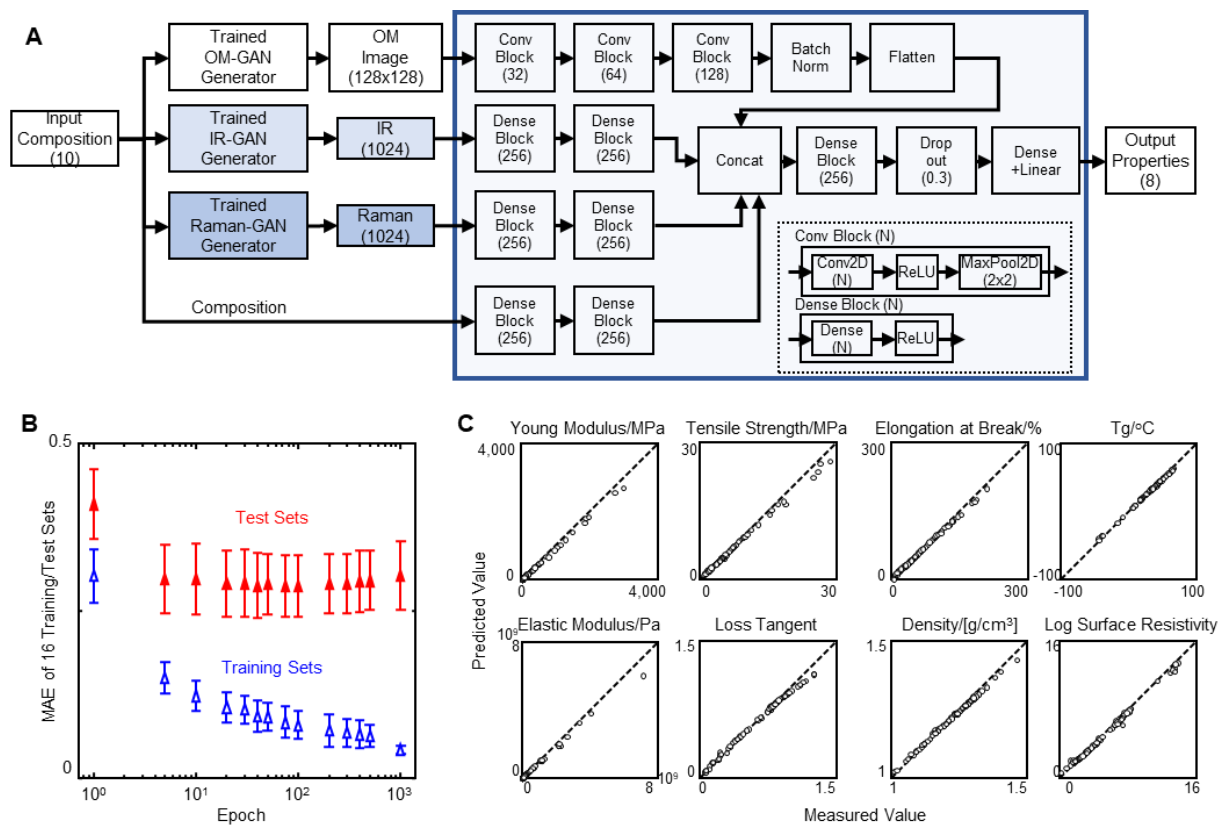

**Fig. 3. Multimodal Deep Learning Predictions of Various Properties.** (A) A diagram of the multimodal deep learning with inputs of optical microscope images, infrared spectra, Raman spectra, and compositional information. (B) Optimization of the multimodal model using mean absolute errors (MAE) averaged across 16 different training/test sets (with a fixed ratio of 7:3 for training and test) to eliminate bias in the selection of samples in the training set. (C) Comparison of predicted properties using the final fully trained model with measured properties.

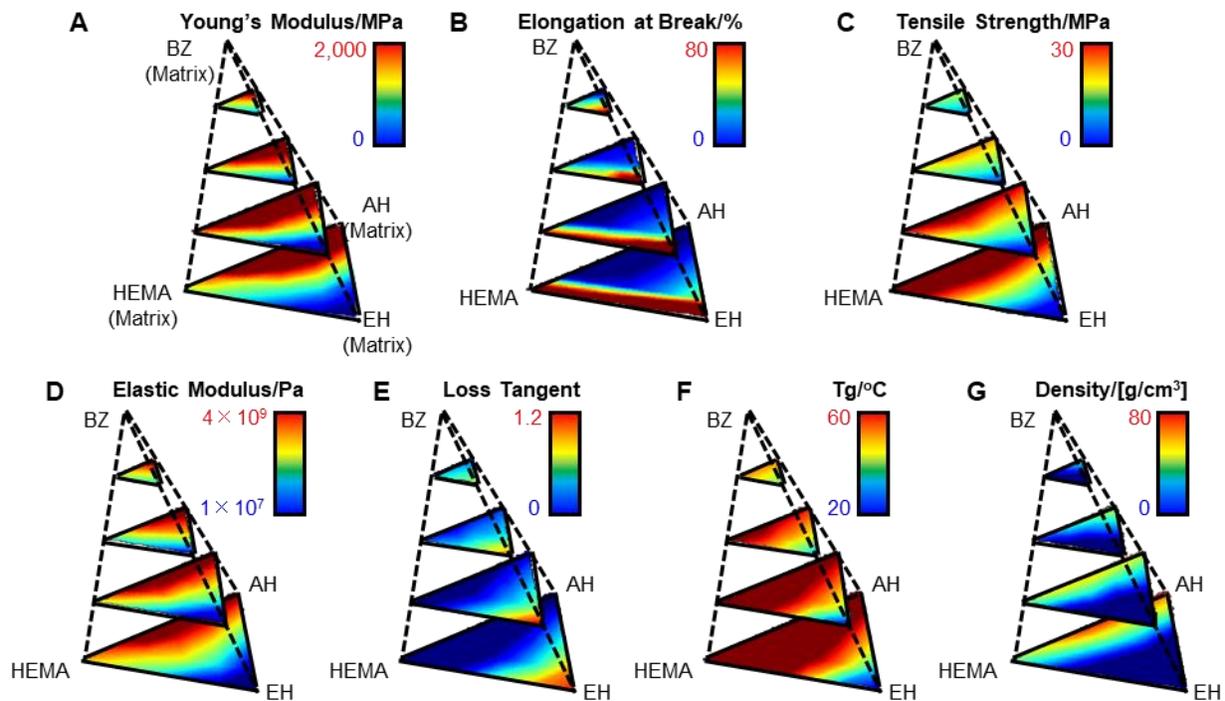

**Fig. 4. Visualization of Multi-dimensional Predictions.** The triangular pyramid diagrams display the multimodal deep learning predictions of (A) Young's modulus, (B) elongation at break, (C) tensile strength, (D) elastic modulus, (E) loss tangent, (F) glass transition temperature (Tg), and (G) density for different compositions of four matrices: AH, EH, HEMA, and BZ.

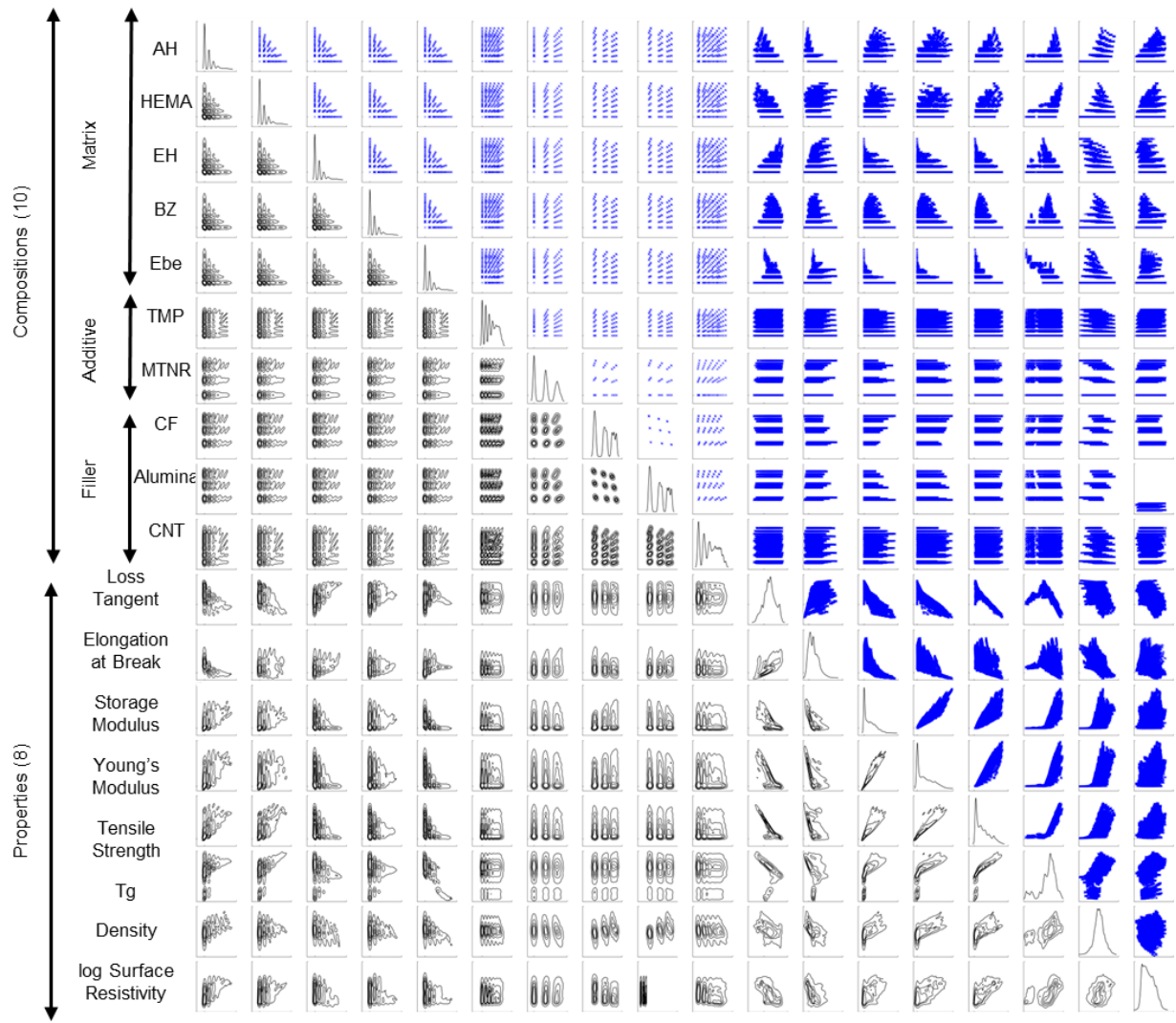

**Fig. 5. A Vast Number of Predictions Encompassing High-dimensional Combinations.** The figures present a total of 17,474,130 data points in the form of pair-plots (Ashby-plots) and contour plots, generated from 114,210 predictions made using our proposed multimodal deep learning. The upper-right figures depict scatter plots, while the lower-left figures illustrate contours of kernel density, obtained from all predictions.

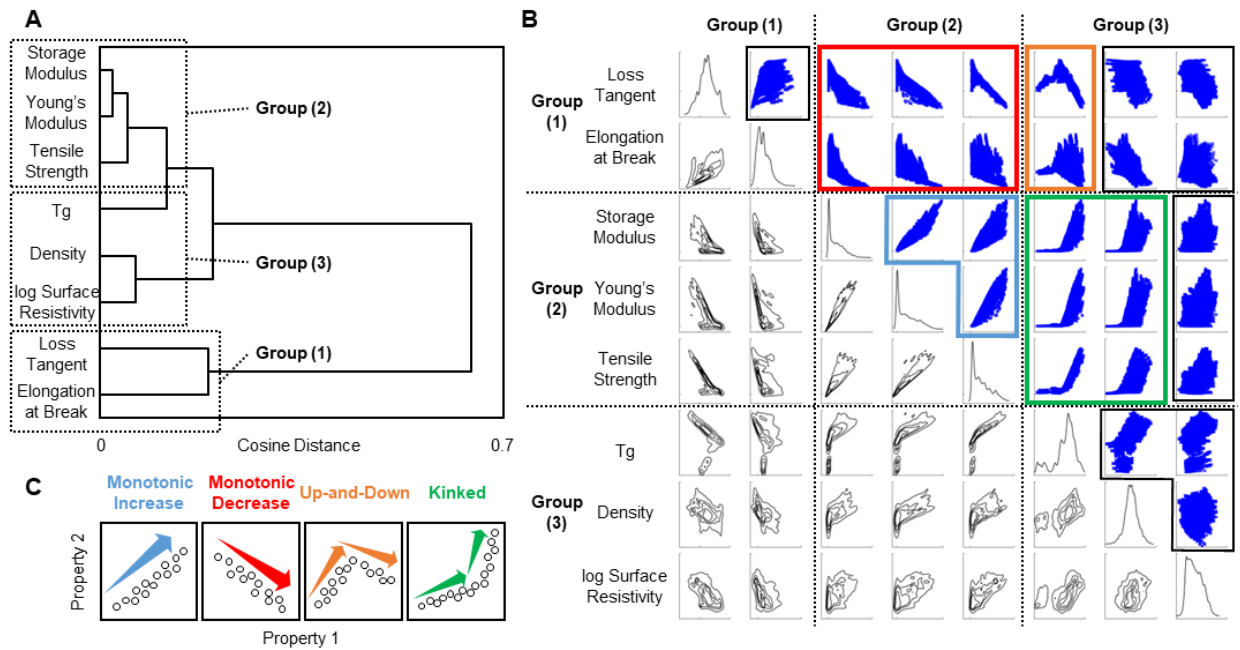

**Fig. 6 Characterization of High-dimensional Predictions.** (A) Dendrogram showing similarities between eight physical properties of 114,210 compositions predicted by multimodal deep learning. (B) Ashby-plots pf physical properties coordinated by the three groups identified from the dendrogram. (C) Typical four trends of Ashby-plots.

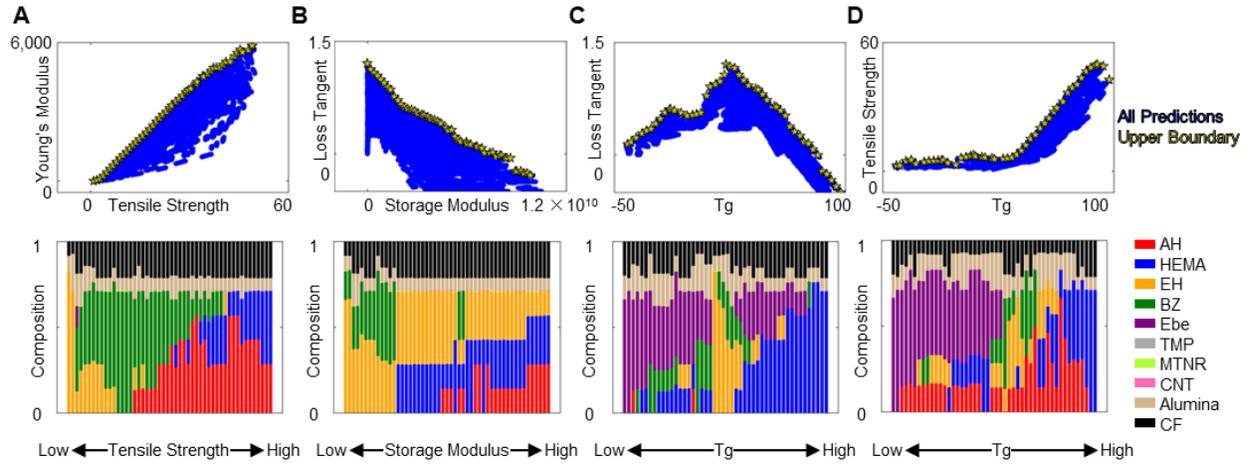

**Fig. 7 Inverse Materials Design from High-dimensional Predictions.** The figures show the distribution of (A) monotonic increase, (B) monotonic decrease (trade-off), (C) up-and-down, (D) kinked relationships between different properties, as predicted by the multimodal deep learning, using 114,210 predictions. The top figures depict scatter plots of the properties, and the bottom figures show stacked bars representing the compositions at the upper boundaries of scatter plots, as indicated by the yellow markers in the corresponding top figures. These relationships include (A) Young's modulus vs tensile strength, (B) loss tangent vs storage modulus, (C) loss tangent vs glass transition temperature (Tg), and (D) tensile strength vs Tg.

# Supplementary Materials for

## A Comprehensive and Versatile Multimodal Deep Learning Approach for Predicting Diverse Properties of Advanced Materials

Shun Muroga[1], Yasuaki Miki[1], and Kenji Hata[1*]

[1]: Nano Carbon Device Research Center, National Institute of Advanced Industrial Science and Technology, Tsukuba Central 5, 1-1-1, Higashi, Tsukuba, Ibaraki, 305-8565, Japan

[*]Corresponding author. E-mail: kenji-hata@aist.go.jp

**This PDF file includes:**

Figs. S1 to S10

Table S1

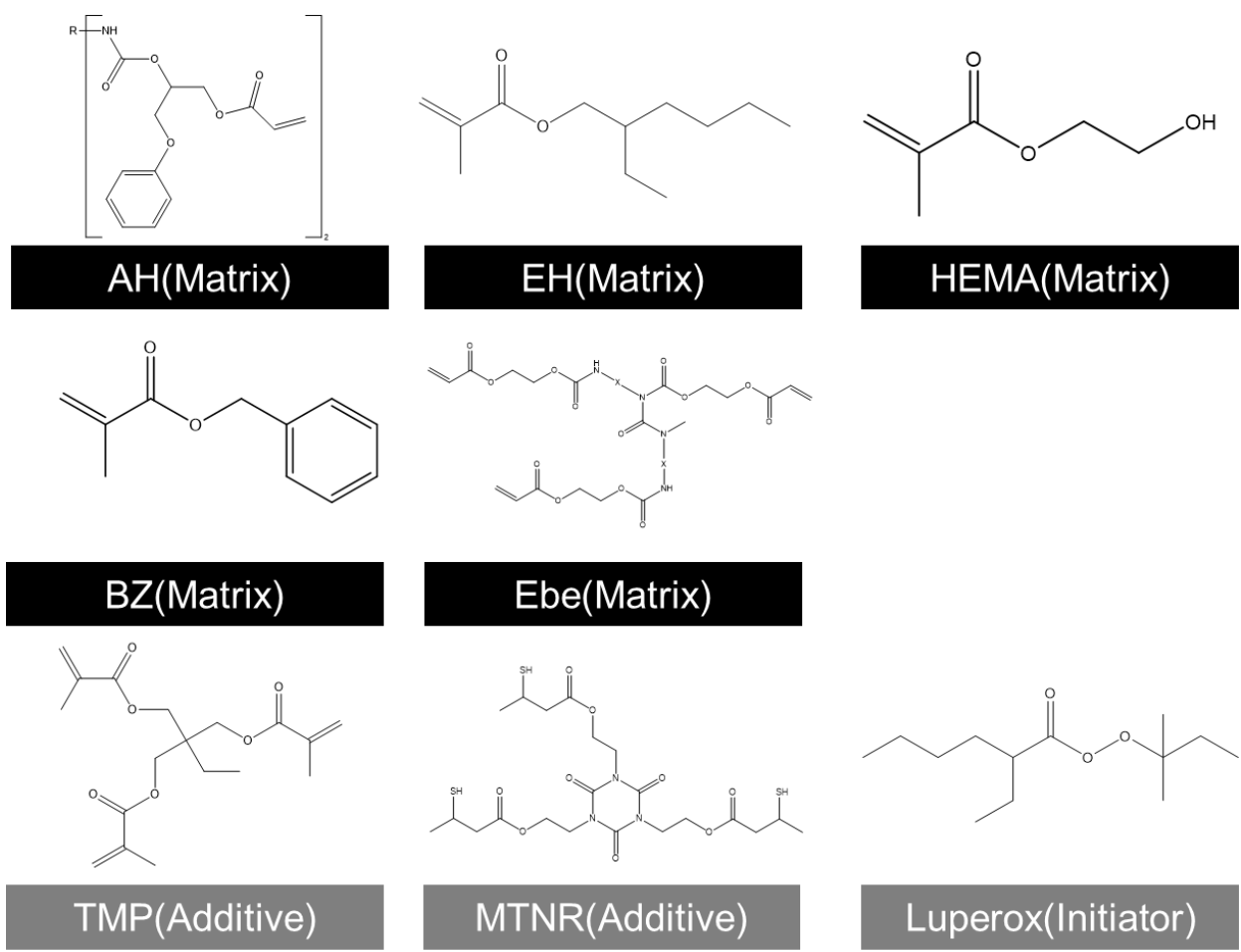

**Fig. S1.** Schematics of molecular structures of matrix monomers and additives used in this study.

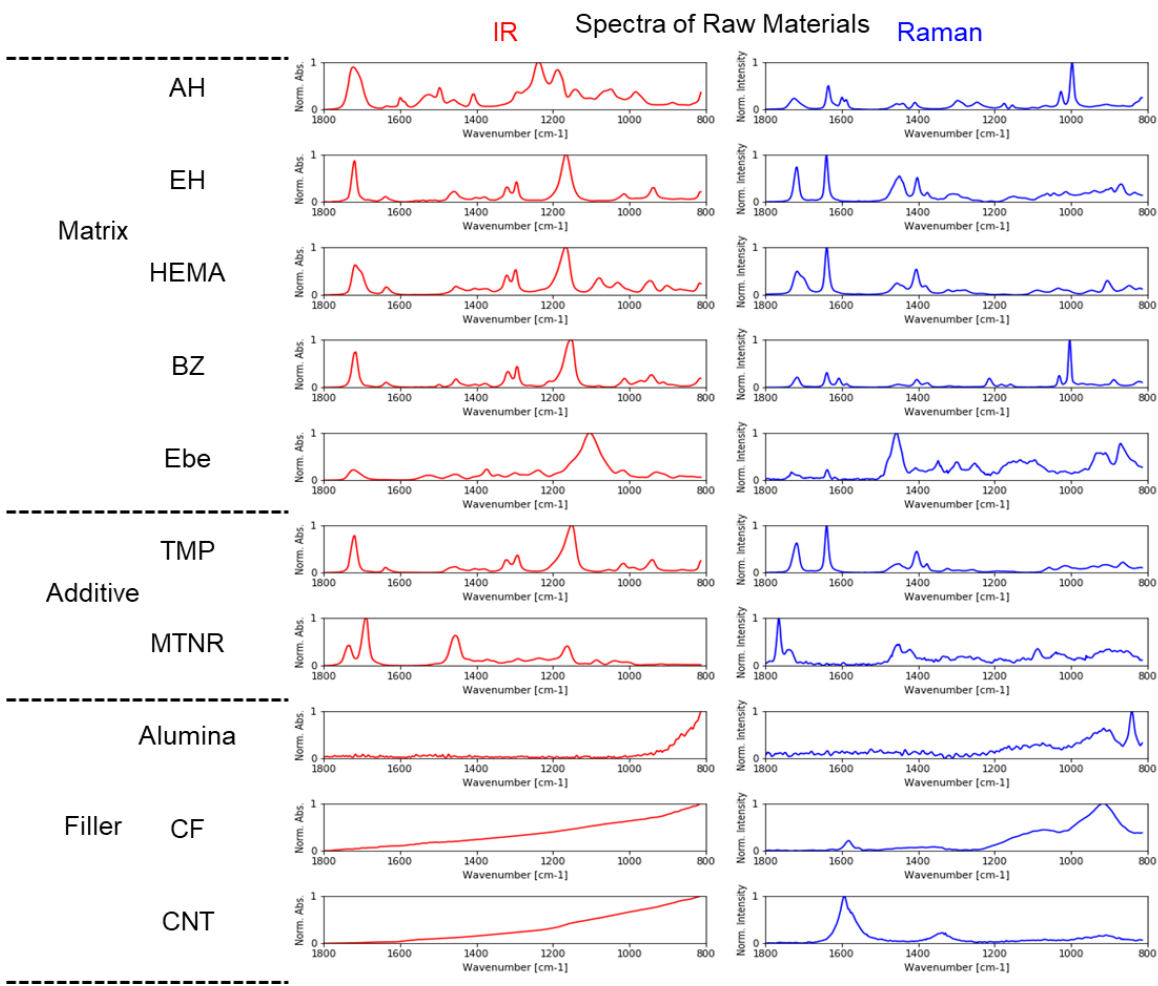

**Fig. S2.** IR and Raman spectra of individual raw matrix monomers, additives, and fillers without any dispersion and curing.

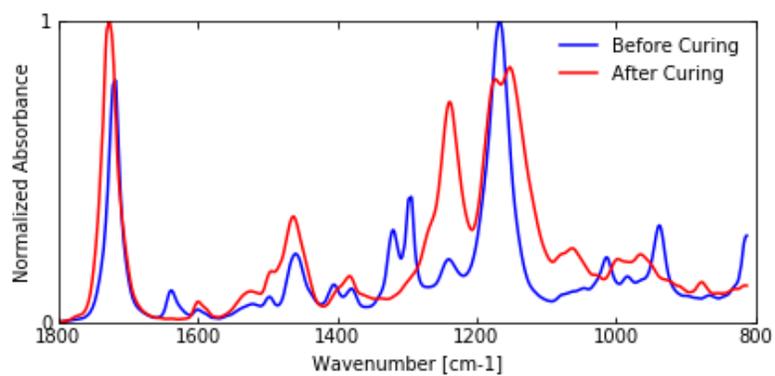

**Fig. S3.** IR spectra of the polymer composite before and after curing. The composition of the polymer composite is AH20/EH80/CF20/Alumina20/CNT0.1.

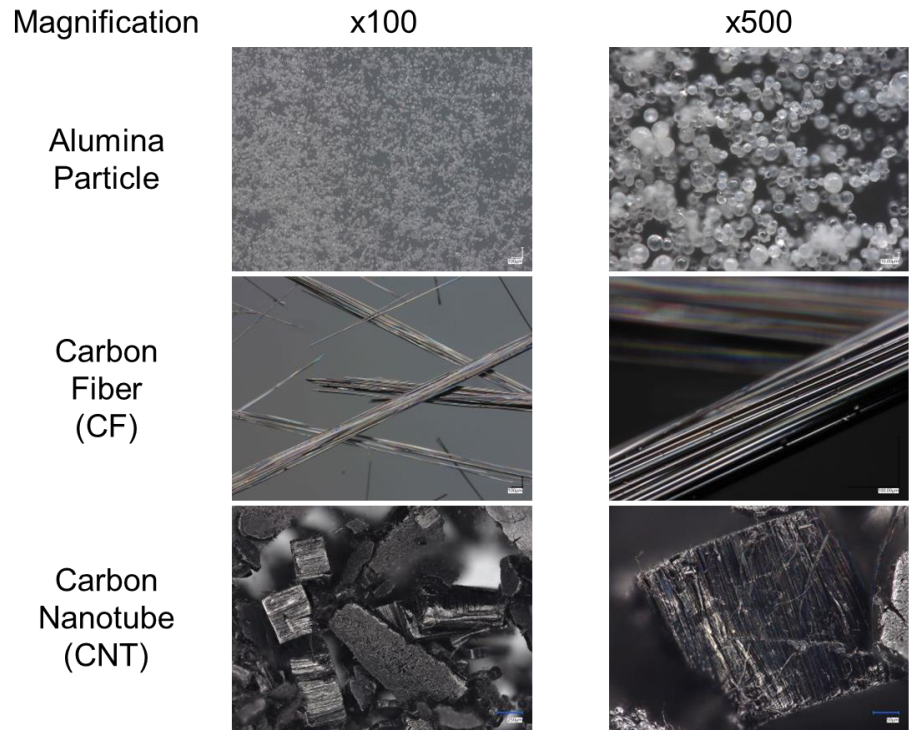

**Fig. S4.** Optical microscope images of raw fillers without any dispersion and curing.

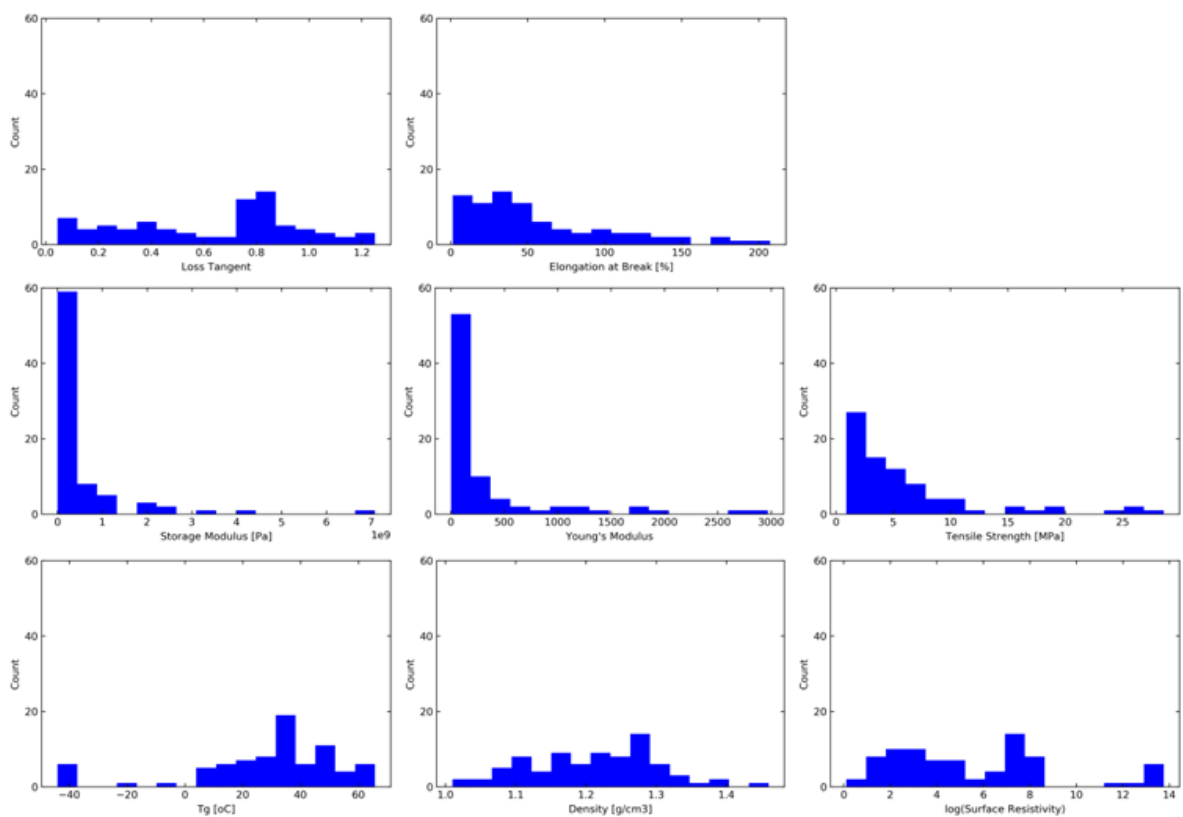

**Fig. S5.** Distributions of the eight physical properties of the 80 polymer composites.

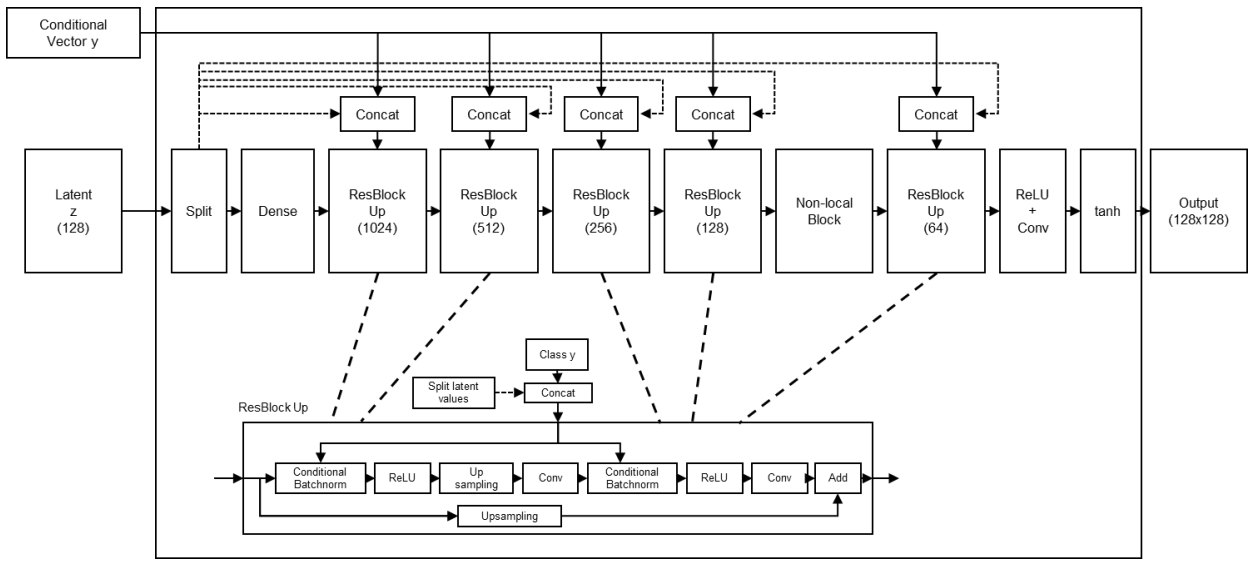

**Fig. S6.** Architecture of BigGAN generator (128x128) for OM-GAN in this study.

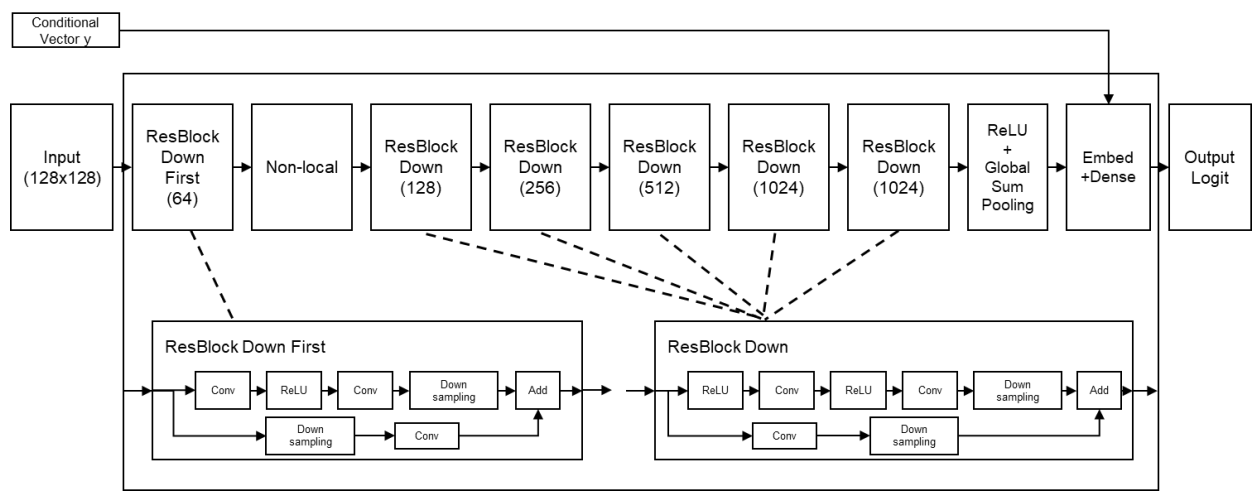

**Fig. S7.** Architecture of BigGAN discriminator (128x128) for OM-GAN in this study.

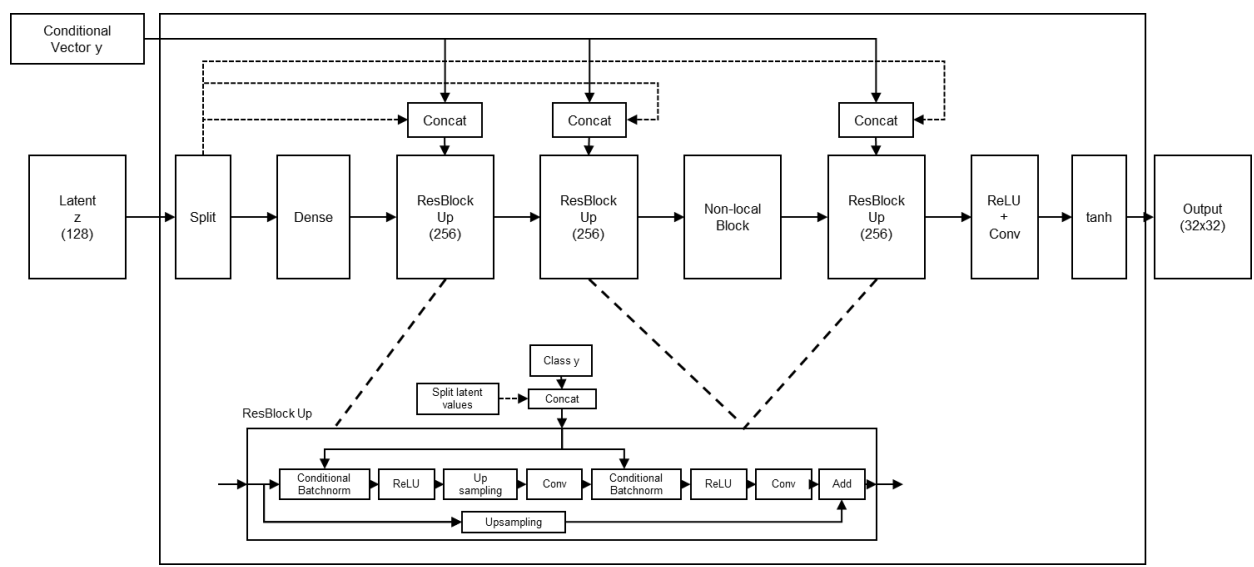

**Fig. S8.** Architecture of BigGAN generator (32x32) for IR-GAN and Raman-GAN in this study.

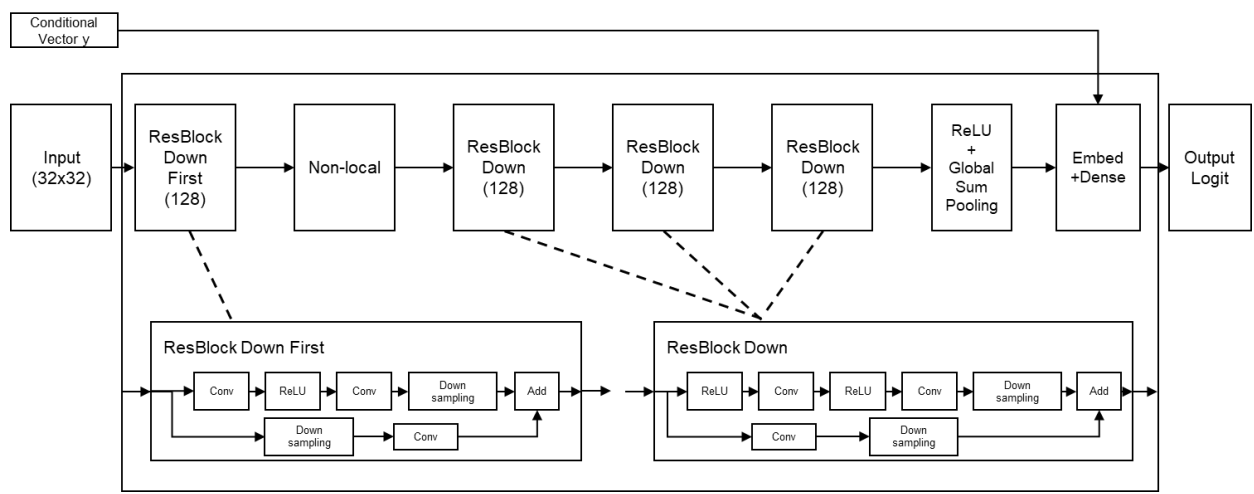

**Fig. S9.** Architecture of BigGAN discriminator (32x32) for IR-GAN and Raman-GAN in this study.

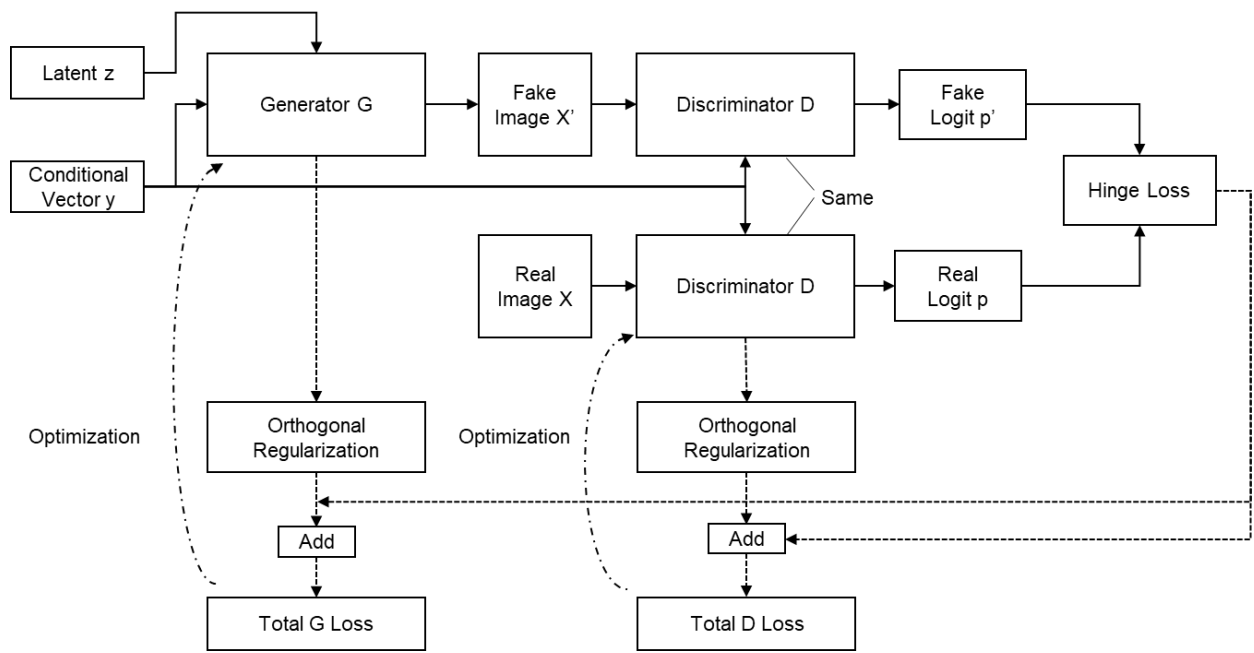

**Fig. S10.** Training protocol of BigGAN generator and discriminator.

**Table S1.** Compositional information of 80 polymer composite samples used in this study.

| AH | EH | HEMA | BZ | Ebe | TMP | MTNR | CNT | Alumina | CF |
|---|---|---|---|---|---|---|---|---|---|
| 0 | 80 | 0 | 0 | 20 | 0 | 0 | 0.2 | 0 | 10 |
| 0 | 80 | 0 | 0 | 20 | 0 | 0 | 0.2 | 0 | 0 |
| 0 | 80 | 0 | 0 | 20 | 0 | 0 | 0.2 | 0 | 20 |
| 0 | 60 | 0 | 0 | 40 | 0 | 0 | 0.1 | 20 | 0 |
| 0 | 41 | 0 | 31 | 28 | 0 | 0 | 0.1 | 10 | 0 |
| 0 | 42 | 0 | 30 | 28 | 0 | 0 | 0.1 | 10 | 10 |
| 0 | 40 | 0 | 0 | 60 | 0 | 0 | 0.5 | 0 | 0 |
| 0 | 80 | 0 | 0 | 20 | 0 | 0 | 0.1 | 10 | 10 |
| 0 | 80 | 0 | 0 | 20 | 0 | 0 | 0.2 | 10 | 10 |
| 0 | 42 | 0 | 30 | 28 | 0 | 0 | 0.1 | 0 | 10 |
| 0 | 60 | 0 | 0 | 40 | 0 | 0 | 0.1 | 10 | 10 |
| 0 | 60 | 0 | 0 | 40 | 0 | 0 | 0.5 | 10 | 10 |
| 0 | 42 | 0 | 31 | 28 | 0 | 0 | 0.2 | 10 | 10 |
| 0 | 60 | 0 | 0 | 40 | 0 | 0 | 0.1 | 20 | 20 |
| 0 | 80 | 0 | 0 | 20 | 0 | 0 | 0.2 | 20 | 20 |
| 0 | 60 | 0 | 0 | 40 | 0 | 0 | 0.5 | 20 | 20 |
| 0 | 56 | 0 | 30 | 14 | 0 | 0 | 0.1 | 10 | 10 |
| 0 | 42 | 0 | 30 | 28 | 0 | 0 | 0.5 | 20 | 20 |
| 0 | 0 | 0 | 0 | 100 | 0 | 0 | 0.1 | 10 | 10 |
| 0 | 80 | 0 | 0 | 20 | 0 | 0 | 0.5 | 20 | 20 |
| 0 | 56 | 0 | 30 | 14 | 0 | 0 | 0.5 | 10 | 10 |
| 0 | 0 | 0 | 0 | 100 | 0 | 0 | 0 | 0 | 20 |
| 0 | 40 | 0 | 0 | 60 | 0 | 0 | 0 | 0 | 20 |
| 0 | 56 | 0 | 30 | 14 | 0 | 0 | 0.2 | 20 | 20 |
| 0 | 42 | 15 | 15 | 28 | 0 | 0 | 0.5 | 20 | 20 |
| 0 | 0 | 0 | 0 | 100 | 0 | 0 | 0.5 | 20 | 20 |
| 0 | 20 | 0 | 0 | 80 | 0 | 0 | 0.5 | 20 | 20 |
| 0 | 52 | 0 | 30 | 17 | 0 | 1 | 0.5 | 20 | 20 |
| 0 | 56 | 0 | 30 | 14 | 0 | 0 | 0.5 | 20 | 20 |
| 20 | 0 | 0 | 0 | 80 | 0 | 0 | 0.1 | 20 | 20 |
| 20 | 0 | 0 | 0 | 80 | 0 | 0 | 0.5 | 20 | 20 |
| 0 | 52 | 0 | 30 | 17 | 1 | 0 | 0.5 | 20 | 20 |

| 0  | 52 | 15 | 15 | 17 | 0 | 1 | 0.5 | 20 | 20 |
|----|----|----|----|----|---|---|-----|----|----|
| 20 | 80 | 0  | 0  | 0  | 0 | 0 | 0.3 | 20 | 0  |
| 20 | 80 | 0  | 0  | 0  | 0 | 0 | 0.1 | 5  | 5  |
| 20 | 80 | 0  | 0  | 0  | 0 | 0 | 0.1 | 10 | 5  |
| 20 | 80 | 0  | 0  | 0  | 0 | 0 | 0.1 | 10 | 10 |
| 0  | 56 | 30 | 0  | 14 | 0 | 0 | 0.2 | 10 | 10 |
| 20 | 80 | 0  | 0  | 0  | 0 | 0 | 0.1 | 15 | 10 |
| 0  | 52 | 30 | 0  | 17 | 0 | 1 | 0.5 | 20 | 20 |
| 20 | 80 | 0  | 0  | 0  | 0 | 0 | 0.1 | 5  | 10 |
| 20 | 80 | 0  | 0  | 0  | 0 | 0 | 0.1 | 20 | 10 |
| 20 | 80 | 0  | 0  | 0  | 0 | 0 | 0.3 | 20 | 20 |
| 20 | 80 | 0  | 0  | 0  | 0 | 0 | 0.5 | 20 | 10 |
| 20 | 80 | 0  | 0  | 0  | 0 | 0 | 0.1 | 10 | 15 |
| 0  | 52 | 15 | 15 | 17 | 1 | 0 | 0.5 | 20 | 20 |
| 20 | 80 | 0  | 0  | 0  | 0 | 0 | 0.5 | 10 | 10 |
| 20 | 80 | 0  | 0  | 0  | 0 | 0 | 0.1 | 10 | 20 |
| 20 | 80 | 0  | 0  | 0  | 0 | 1 | 0.5 | 20 | 20 |
| 40 | 0  | 0  | 0  | 60 | 0 | 0 | 0.1 | 10 | 10 |
| 20 | 80 | 0  | 0  | 0  | 0 | 0 | 0.5 | 20 | 20 |
| 20 | 80 | 0  | 0  | 0  | 0 | 0 | 0.5 | 10 | 20 |
| 21 | 79 | 0  | 0  | 0  | 0 | 1 | 0.5 | 0  | 20 |
| 40 | 0  | 0  | 0  | 60 | 0 | 0 | 0.1 | 20 | 20 |
| 18 | 53 | 0  | 30 | 0  | 0 | 0 | 0.1 | 20 | 0  |
| 18 | 53 | 0  | 30 | 0  | 0 | 0 | 0.3 | 20 | 0  |
| 0  | 52 | 30 | 0  | 17 | 1 | 0 | 0.5 | 20 | 20 |
| 14 | 56 | 15 | 15 | 0  | 0 | 1 | 0.1 | 20 | 0  |
| 20 | 80 | 0  | 0  | 0  | 1 | 0 | 0.5 | 20 | 20 |
| 20 | 80 | 0  | 0  | 0  | 0 | 0 | 0.5 | 0  | 20 |
| 20 | 80 | 0  | 0  | 0  | 1 | 0 | 0.5 | 0  | 20 |
| 0  | 52 | 0  | 30 | 17 | 5 | 0 | 0.5 | 20 | 20 |
| 20 | 80 | 0  | 0  | 0  | 0 | 0 | 0.5 | 30 | 30 |
| 14 | 56 | 15 | 15 | 0  | 0 | 1 | 0.1 | 0  | 20 |
| 14 | 56 | 15 | 15 | 0  | 0 | 1 | 0.3 | 0  | 20 |
| 14 | 56 | 15 | 15 | 0  | 1 | 0 | 0.1 | 20 | 0  |
| 0  | 52 | 15 | 15 | 17 | 5 | 0 | 0.5 | 20 | 20 |
| 14 | 56 | 15 | 15 | 0  | 1 | 0 | 0.3 | 20 | 0  |
| 18 | 52 | 30 | 0 | 0 | 0 | 0 | 0.1 | 10 | 10 |
| 0 | 53 | 30 | 0 | 17 | 5 | 0 | 0.5 | 20 | 20 |
| 20 | 80 | 0 | 0 | 0 | 5 | 0 | 0.5 | 20 | 20 |
| 20 | 80 | 0 | 0 | 0 | 5 | 0 | 0.5 | 0 | 20 |
| 60 | 0 | 0 | 0 | 40 | 0 | 0 | 0.1 | 10 | 10 |
| 14 | 56 | 15 | 15 | 0 | 1 | 0 | 0.1 | 0 | 20 |
| 60 | 0 | 0 | 0 | 40 | 0 | 0 | 0.5 | 20 | 20 |
| 18 | 52 | 30 | 0 | 0 | 0 | 0 | 0.1 | 20 | 10 |
| 17 | 52 | 0 | 30 | 0 | 0 | 0 | 0.3 | 0 | 20 |
| 80 | 0 | 0 | 0 | 20 | 0 | 0 | 0.1 | 10 | 10 |
| 80 | 0 | 0 | 0 | 20 | 0 | 0 | 0.5 | 20 | 20 |
| 17 | 53 | 30 | 0 | 0 | 0 | 0 | 0.1 | 10 | 20 |